\documentclass[aps,pra,preprint,amsmath]{revtex4-1}
\usepackage{silence}
\usepackage{mathrsfs}
\usepackage{graphicx}
\usepackage{epstopdf}
\usepackage{amsthm}
\usepackage{amsmath}
\usepackage{amsfonts}
\usepackage{amssymb}
\usepackage{braket}
\usepackage{bbold}
\usepackage{subfigure}
\usepackage{appendix}
\usepackage{tikz}
\usepackage{bm}
\usepackage{float}
\def\be{\begin{equation}}
\def\ee{\end{equation}}
\def\ba{\begin{array}}
\def\ea{\end{array}}

\def\1{{\bf{1}}}

\input amssym.def
\begin{document}
\title{\bf Uncertainty relations based on the $\rho$-absolute variance for quantum channels}

\author{Cong Xu$^1$}
\author{Wen Zhou$^1$}
\author{Qing-Hua Zhang$^2$}
\author{Shao-Ming Fei$^{1}$}
\email{Cong Xu\\2524596282@qq.com\\Shao-Ming Fei\\feishm@cnu.edu.cn}
\affiliation{
$^1$School of Mathematical Sciences, Capital Normal University, Beijing 100048, China\\
$^2$School of Mathematics and Statistics, Changsha University of Science and Technology, Changsha 410114, China
}

\begin{abstract}
Uncertainty principle reveals the intrinsic differences between the classical and quantum worlds, which plays a significant role in quantum information theory. By using $\rho$-absolute variance, we introduce the uncertainty of quantum channels and explore its properties. By using Cauchy-Schwarz inequality and the parallelogram law, we establish the product and summation forms of the uncertainty relations for arbitrary two quantum channels, respectively. The summation form of the uncertainty inequalities based on the $\rho$-absolute variance for arbitrary $N$ quantum channels are also investigated and the optimal lower bounds are presented. We illustrate our results by several typical examples.

\end{abstract}

\maketitle

\noindent {\bf 1. Introduction}

As one of the fundamental building blocks of quantum mechanics, uncertainty principle has wide applications in quantum cryptography \cite{FCP,RJB}, entanglement detection \cite{BSL,GO,HJBR}, quantum nonlocality \cite{OW} and so on. Heisenberg introduced originally the uncertainty principle in $1927$ \cite{HW}. The well-known uncertainty principle was given by Weyl and Kennard \cite{WH,KEH} in terms of the variances of position and momentum. The uncertainty principle implies that one can not measure precisely the position and momentum of a quantum system simultaneously. Robertson \cite{RH} generalized the uncertainty principle for position and momentum to arbitrary two observables $A$ and $B$ with respect to a quantum state $|\psi\rangle$,
\begin{align}\label{eq1}
 V_{|\psi\rangle}(A) V_{|\psi\rangle}(B)\geq \left|\frac{1}{2}\langle\psi|[A,B]|\psi\rangle\right|^2,
\end{align}
where $[A,B]=AB-BA$ and $V_{|\psi\rangle}(M)=\langle\psi| M^2|\psi\rangle-\langle\psi| M|\psi\rangle^2$.  Subsequently, the inequality (\ref{eq1}) was improved by Schr\"{o}dinger \cite{SE},
\begin{align}\label{eq2}
 V_{|\psi\rangle}(A) V_{|\psi\rangle}(B)\geq  \left|\frac{1}{2}\langle[A,B]\rangle\right|^2+\left|\frac{1}{2}\langle\{A,B\}\rangle-\langle A\rangle\langle B\rangle\right|^2,
\end{align}
where $\{A,B\}=AB+BA$ and $\langle M\rangle=\langle\psi|M|\psi\rangle$. The lower bounds of the inequalities $(\ref{eq1})$ and $(\ref{eq2})$ become trivial if the state $|\psi\rangle$ is an eigenstate of the observable $A$ or $B$. Later, Maccone $et \ al.$ \cite{MP} presented the following two uncertainty relations based on the sum of variances,
\begin{align}\label{eq3}
V_{|\psi\rangle}(A)+V_{|\psi\rangle}(B)\geq\pm i \langle\psi|[A,B]|\psi\rangle+|\langle\psi|A\pm iB|\psi^\perp\rangle|^2
\end{align}
and
\begin{align}\label{eq4}
V_{|\psi\rangle}(A)+V_{|\psi\rangle}(B)\geq\frac{1}{2}|\langle\psi^\perp_{A+B}|A+B|\psi\rangle|^2.
\end{align}
The inequality (\ref{eq3}) ((\ref{eq4})) is valid for arbitrary states $|\psi^\perp\rangle$ ($|\psi^\perp_{A+B}\rangle\propto(A+B-\langle A+B\rangle)|\psi\rangle$) orthogonal to the system state $|\psi\rangle$. Both uncertainty relations have nontrivial lower bounds and have been tested experimentally \cite{WZB}.


As a mixture of classical uncertainty and quantum uncertainty \cite{LS1,LS2,LS3}, variance quantifies the total uncertainty of the observable in the quantum state $\rho$. By removing the restriction on operators that are Hermitian, the $\rho$-absolute variance of arbitrary operator $K$ was introduced by Gudder \cite{Gudder}. In Refs. \cite{DD1,SL} the authors defined another non-Hermitian extension of the variance, the so called modified variance. Wu $et \  al.$ \cite{Wu1} generalized the modified variance to the two parameter version.
By considering state-channel interaction, Luo $et \ al.$ \cite{LS} defined the quantity $I_{\rho}(\Phi)$ as the quantum uncertainty of the quantum channel $\Phi$ in the quantum state $\rho$. Sun $et \ al.$ \cite{SL} explored the quantum/total uncertainty of quantum channels based on the modified skew information/variance in which the operator are not necessarily Hermitian. Xu $et \ al.$ \cite{XWF3} generalized the results in Ref. \cite{SL} to a more general case.

Analogous to the Heisenberg's uncertainty principle, the product form of the uncertainty relations for quantum channels reveals the quantum channels' fundamental properties. By using the Cauchy-Schwarz inequality, Zhou $et \ al.$ \cite{ZN} introduced the product form of uncertainty relations for two quantum channels based on the modified skew information. Recently, the product and summation forms of the uncertainty relations for quantum channels based on the variance and skew information have also been studied intensely \cite{XWF2,XWF1,XWF4,ZL,ZWF1,ZWF2,ZWF5,ZWF7,FSS,CAL,RRNL,HLTG,HJ1,ZL}.

The remainder of this paper is structured as follows. In Section 2 we introduce the uncertainty relations for quantum channel $\Phi$ based on the $\rho$-absolute variance and prove that it satisfies several properties. We then establish the product and summation forms of the uncertainty relations for two quantum channels. The summation form of the uncertainty relations for arbitrary $N$ quantum channels in terms of the $\rho$-absolute variance are explored in Section 3. Finally we conclude with a summary in Section 4.

\noindent {\bf 2. Uncertainty relations for two quantum channels}

Let $\mathcal{H}$ be an $n$-dimensional Hilbert space. Denote
$\mathcal{B(H)}$, $\mathcal{S(H)}$ and $\mathcal{D(H)}$ the set of
all bounded linear operators, Hermitian operators and density
operators (positive operators with trace one) on $\mathcal{H}$,
respectively. The expectation value of an operator $A\in\mathcal{S(H)}$ with respect to a quantum state $\rho$ is $E_{\rho}(A)=\mathrm{Tr}(\rho A)$.
For $X,Y\in\mathcal{B(H)}$, $\langle X,Y\rangle=\mathrm{Tr}(X^\dag Y)$ is the inner product. The norm of $X\in\mathcal{B(H)}$ is defined by $\|X\|=\mathrm{Tr}(X^{\dag}X)^\frac{1}{2}$.

For a quantum state $\rho\in\mathcal{D(H)}$ and an observable $A\in\mathcal{S(H)}$, the $\rho$-variance of $A$ is defined by \cite{Gudder}
\begin{align}\label{eq5}
 V_{\rho}(A)=E_{\rho}(A^2_0)=E_{\rho}(A^2)-E_{\rho}(A)^2,
\end{align}
where $A_0=A-\mathrm{Tr}(\rho A)$.
We denote $|K|=(K^{\dag}K)^\frac{1}{2}\in\mathcal{S(H)}$ for $K\in\mathcal{B(H)}$. The $\rho$-absolute variance of $K$ is defined by \cite{Gudder}
\begin{align}\label{eq6}
 |V_{\rho}|(K):=E_{\rho}(|K_0|^2)=\|K_0\sqrt{\rho}\|^{2}=E_{\rho}(|K|^2)-|E_{\rho}(K)|^2,
\end{align}
where $K_0=K-\mathrm{Tr}(\rho K)$.

Consider a quantum channel $\Phi$ with Kraus representation
$\Phi(\rho)=\sum_{i}K_i\rho K_i^\dag$, where $\sum_{i}K_i^\dag K_i=I$ with $I$ the identity operator. We introduce the uncertainty of the quantum channel $\Phi$ based on the $\rho$-absolute variance,
\begin{align}\label{eq7}
   |V_{\rho}|(\Phi)=\sum_{i}|V_{\rho}|(K_i)=&\sum_{i}\|\widetilde{K}_i\|^{2}  \notag\\
   =&1-\sum_{i}|\mathrm{Tr}(\rho K_i)|^2,
\end{align}
where $\widetilde{K}_i=K_{i0}\sqrt{\rho}$ with $K_{i0}=K_i-\mathrm{Tr}(\rho K_i)$.

The Kraus representations of a channel are usually not unique. Two sets of Kraus operators $\{K_i\}$ and $\{L_j\}$ describe the
same quantum channel $\Phi$ if and only if there exists a unitary matrix
${U=(u_{ij})}$ such that $K_i=\sum_j u_{ij} L_j$ for any $i$
\cite{NC}. From equation (\ref{eq7}) it is easily verified that $|V_{\rho}|(\Phi)$ is independent of the choice of the
Kraus operators of $\Phi$, namely, $|V_{\rho}|(\Phi)$ is well-defined.
We can prove that the uncertainty $|V_{\rho}|(\Phi)$ has the following properties:

(i) (Non-negativity) $|V_{\rho}|(\Phi)\geq 0$, with the
equality holds if and only if $K_i\sqrt{\rho}=(\mathrm{Tr}\rho K_i)\sqrt{\rho}$ for any $i$.

(ii) (Linearity) For any $\lambda_1,\lambda_2 \geq
0$ and any quantum channels $\Phi_1$ and $\Phi_2$,
$|V_{\rho}|(\lambda_1\Phi_1+\lambda_2\Phi_2)=\lambda_1
|V_{\rho}|(\Phi_1)+\lambda_2
|V_{\rho}|(\Phi_2)$.

(iii) (Concavity) $|V_{\rho}|(\Phi)$ is concave in $\rho$, i.e.,
$|V_{\sum_j\lambda_j\rho_j}|(\Phi)\geq
\sum_j\lambda_j|V_{\rho_j}|(\Phi)$, where $\lambda_j\geq
0$ for each $j$ with $\sum_j\lambda_j=1$.

(iv) (Unitary invariance) $|V_{U\rho U^\dag}|(U\Phi
U^\dag)=|V_{\rho}|(\Phi)$ for any unitary operators
$U$, where $U\Phi U^\dag(\rho)=\sum_i(U K_iU^\dag)$ $\rho (U K_i
U^\dag)^\dag$ with $\Phi(\rho)=\sum_{i}K_i\rho K_i^\dag$.

(v) (Ancillary independence)
$|V_{\rho^{ab}}|(\Phi^a\otimes\mathcal{I}^b)=|V_{\rho^a}|(\Phi^a)$,
where $\rho^{ab}$ is any bipartite state share by two parties $a$ and
$b$, $\rho^{a}=\mathrm{Tr}_b
[\rho^{ab}]$ is the reduced state on party $a$, and $\mathcal{I}^b$ is the identity channel on
system $b$.

(vi) (Additivity)
$|V_{\rho^{ab}}|(\Phi^a\otimes\mathcal{I}^b+\mathcal{I}^a\otimes\Phi^b)=
|V_{\rho^a}|(\Phi^a)+|V_{\rho^b}|(\Phi^b)$,
where ${\rho^a}=\mathrm{Tr}_b[\rho^{ab}]$
($\rho^b=\mathrm{Tr}_a[\rho^ {ab}]$)
is the reduced state with respect to the subsystems $a$ ($b$),
$\Phi^a$ ($\Phi^b$) and $\mathcal{I}^a$ ($\mathcal{I}^b$) are the channel and
the identity channel on systems $a$ ($b$), respectively.

The above properties satisfied by $|V_{\rho}|(\Phi)$ can be seen in the following way. By definition, $|V_{\rho}|(\Phi)\geq 0$ is obvious and $|V_{\rho}|(\Phi)=0$ if and only if $\|K_{i0}\sqrt{\rho}\|=0$, i.e., $K_i\sqrt{\rho}=(\mathrm{Tr}\rho K_i)\sqrt{\rho}$.
Let $\Phi_1(\rho)=\sum_i^fK_i\rho K^{\dag}_i$, $\Phi_2(\rho)=\sum_j^mL_j\rho L^{\dag}_j$ and $\widetilde{\Phi}(\rho)=\sum_s^{f+m}F_s\rho F^{\dag}_s$, where
$$ F_k=\left\{
\begin{aligned}
\sqrt{\lambda_1}K_i, &&s=i,&&&&&&s \in\{1,2,\cdots, f\}\\
\sqrt{\lambda_2}L_j, &&s=j,&&&&&&s \in\{f+1,f+2,\cdots, f+m\}
\end{aligned}
\right.
$$
i.e., $\widetilde{\Phi}(\rho)=\lambda_1\Phi_1(\rho)+\lambda_2\Phi_2(\rho)$.
By rewriting (\ref{eq7}) as
\begin{align*}
   |V_{\rho}|(\Phi)=\sum_i\mathrm{Tr}\rho K^{\dag}_iK_i-\sum_i\sum^n_{l,k=1}\langle k|\rho K^{\dag}_i |k\rangle\langle l|\rho K_i|l\rangle,
\end{align*}
where $\{|t\rangle\}^n_{t=1}$ is an orthonormal basis in $\mathcal{H}$, it is easily to see that $|V_{\rho}|(\Phi)$ is positive-real-linear of $\Phi$. Hence, item (ii) holds.
From equation (\ref{eq7}) and Cauchy-Schwarz inequality, item (iii) is obviously true. Item (iv) is clear by straightforward derivation using equation (\ref{eq7}). For item (v), let $W$ be any operator on $\mathcal{H}_{a}\otimes\mathcal{H}_{b}$. Then
$\mathrm{Tr}((F_a\otimes I_b)W)=\mathrm{Tr}(F_a\cdot
\mathrm{Tr}_bW)$ for any $F_a$ on $\mathcal{H}_{a}$ \cite{BHATIA}. Item (vi) is obtained from item (ii) and (v).

Next we explore the product and summation forms of uncertainty relations in terms of the $\rho$-absolute variance for two quantum channels.

For any operator $X\in\mathcal{B(H)}$, we write $| X\rangle=(x_{11},\cdots, x_{n1},\cdots,x_{1n},\cdots x_{nn})^T$ under the orthonormal basis $\{|l\rangle\langle k|\}_{l,k=1}^{n}$. Denote $|X^\downarrow\rangle=(x'_{11},x'_{21},\cdots,x'_{nn})^T$ the rearranged matrix of $|X\rangle$ in non-increasing order, $|\langle k| X^\downarrow|l\rangle|\geq|\langle k+1|X^\downarrow|l\rangle|\geq|\langle k'| X^\downarrow|l+1\rangle|$ for any $k$, $k'$ and $l$. By relabeling the sub-indices we write
$|X^\downarrow\rangle=(x'_{1},x'_{2},\cdots,x'_{n^2})^T$ such that $x'_i\geq x'_{i+1}$. For arbitrary two non-increasing $n^2$-tuples $a'=(a'_{1},a'_{2},\cdots,a'_{n^2})$ and $b'=(b'_{1},b'_{2},\cdots,b'_{n^2})$ of nonnegative numbers, Xiao $et \ al$. \cite{YNB} presented the following rearrangement inequality,
\begin{align}\label{eq8}
\sum_{i}^{n^2}(a'_ib'_i)\geq\sum_{i}^{n^2}(a'_ib'_{\widetilde{\pi}(i)})\geq\sum_{i}^{n^2}(a'_ib'_{n^2+1-i}), \end{align}
where $\widetilde{\pi}\in S_{n^2}$ is an arbitrary $n^2$-element permutation.

Similarly, for $|X^\downarrow\rangle=(x'_{11},x'_{21},\cdots,x'_{nn})^T$ and
$|Y^\downarrow\rangle=(y'_{11},y'_{21},\cdots,y'_{nn})^T$ we have
\begin{align}\label{eq9}
\sum_{i}^{n^2}|x'_iy'_i|\geq\sum_{i}^{n^2}|x'_iy'_{\widetilde{\pi}(i)}|\geq\sum_{i}^{n^2}|x'_iy'_{n^2+1-i}|. \end{align}
Therefore, we obtain the following product form uncertainty relation.

{\bf Theorem 1} Let $\Phi_1$ and $\Phi_2$ be two quantum
channels on $n$-dimensional Hilbert space $\mathcal{H}$ with Kraus representations
$\Phi_1(\rho)=\sum_{i=1}K_{i}\rho K_i^\dag$ and $\Phi_2(\rho)=\sum_{j=1}L_{j}\rho L_j^\dag$, respectively. The following tight uncertainty relation holds,
\begin{align}\label{eq10}
|V_{\rho}|(\Phi_1)|V_{\rho}|(\Phi_2)\geq\sum_{i,j=1}\left(\sum_{k,l=1}^{n}|\langle l|\widetilde{K}^{\dag}_i|k\rangle'\langle k|\widetilde{L}_j|l\rangle'|\right)^2,
\end{align}
where $'$ means non-increasing order and $\{|t\rangle\}_{t=1}^{n}$ is an
orthonormal basis of the $n$-dimensional Hilbert space $\mathcal{H}$.

{\it Proof} According to Cauchy-Schwarz inequality, we have
\begin{align*}
|V_{\rho}|(K_i)|V_{\rho}|(L_j)=\mathrm{Tr}(\widetilde{K}^\dag_i\widetilde{K}_i)\mathrm{Tr}(\widetilde{L}_j\widetilde{L}^\dag_j)
=&\langle\langle(\widetilde{K}_i)^\downarrow|,|(\widetilde{K}_i)^\downarrow\rangle\rangle\langle\langle(\widetilde{L}_j)^\downarrow|,|(\widetilde{L}_j)^\downarrow\rangle\rangle \notag \\
=&\left(\sum_{k,l=1}^{n}|\langle
l|\widetilde{K}^{\dag}_i|k\rangle'|^2\right)\left(\sum_{\overline{k},\overline{l}=1}^{n}|\langle \overline{k}|\widetilde{L}_j|\overline{l}\rangle'|^2\right) \notag\\
\geq&\left(\sum_{k,l=1}^{n}|\langle l|\widetilde{K}^{\dag}_i|k\rangle'\langle k|\widetilde{L}_j|l\rangle'|\right)^2,
\end{align*}
where $\widetilde{K}_i=K_{i0}\sqrt{\rho}=(K_i-\mathrm{Tr}(\rho K_i))\sqrt{\rho}$. Summing over the index $i,j$, we prove the Theorem 1. $\Box$

%

Similarly, we have the following summation form uncertainty relation.

{\bf Theorem 2} Let $\Phi_1$ and $\Phi_2$ be two quantum
channels on $n$-dimensional Hilbert space $\mathcal{H}$ with Kraus representations
$\Phi_1(\rho)=\sum_{i=1}K_{i}\rho K^\dag_i$ and $\Phi_2(\rho)=\sum_{i=1}L_{i}\rho L^\dag_i$, respectively. Then
\begin{align}\label{eq11}
|V_{\rho}|(\Phi_1)+|V_{\rho}|(\Phi_2)\geq\mathop{\mathrm{max}}\limits_{\pi\in S_n}\frac{1}{2}\sum_{i=1}|V_{\rho}|(K_i\pm L_{\pi(i)}),
\end{align}
where $S_n$ is the $n$-element permutation group and $\pi\in S_n$ is an arbitrary $n$-element permutation.

{\it Proof} Employing the parallelogram law
$\|u+v\|^2+\|u-v\|^2=2(\|u\|^2+\|v\|^2)$, we have
\begin{align*}
|V_{\rho}|({K_i})+|V_{\rho}|({L_{\pi(i)}})
=&(\|\widetilde{K}_i\|^2+\|\widetilde{L}_{\pi (i)}\|^2)\notag\\
=&\frac{1}{2}(\|\widetilde{K}_i+\widetilde{L}_{\pi (i)}\|^2+\|\widetilde{K}_i-\widetilde{L}_{\pi (i)}\|^2)     \notag\\
=& \frac{1}{2}(|V_{\rho}|({K_i+L_{\pi (i)}})+|V_{\rho}|({K_i-L_{\pi (i)}}))\notag\\
\geq&\frac{1}{2}|V_{\rho}|(K_i\pm L_{\pi(i)}),
\end{align*}
where $\widetilde{K}_i=K_{i0}\sqrt{\rho}=(K_i-\mathrm{Tr}(\rho K_i))\sqrt{\rho}$. Consequently, we obtain (\ref{eq11}). $\Box$

In the example below, we illustrate the lower bounds of the product and summation forms uncertainty relations, respectively.

\begin{figure}[ht]\centering
\subfigure[]
{\begin{minipage}[Xu-Cong-uncertainty-6-1a]{0.49\linewidth}
\includegraphics[width=0.98\textwidth]{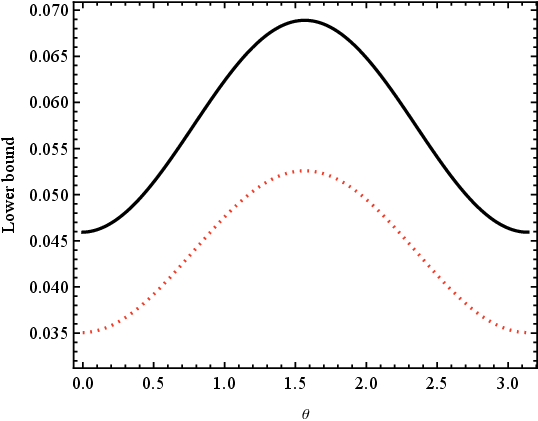}
\end{minipage}}
\subfigure[]
{\begin{minipage}[Xu-Cong-uncertainty-6-1b]{0.49\linewidth}
\includegraphics[width=0.98\textwidth]{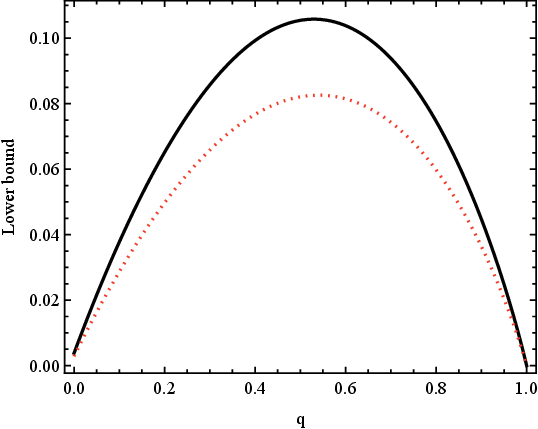}
\end{minipage}}
\subfigure[]
{\begin{minipage}[Xu-Cong-uncertainty-6-1c]{0.49\linewidth}
\includegraphics[width=0.98\textwidth]{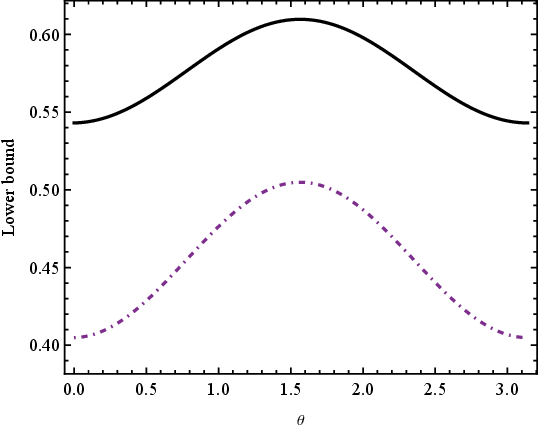}
\end{minipage}}
\subfigure[]
{\begin{minipage}[Xu-Cong-uncertainty-6-1d]{0.49\linewidth}
\includegraphics[width=0.98\textwidth]{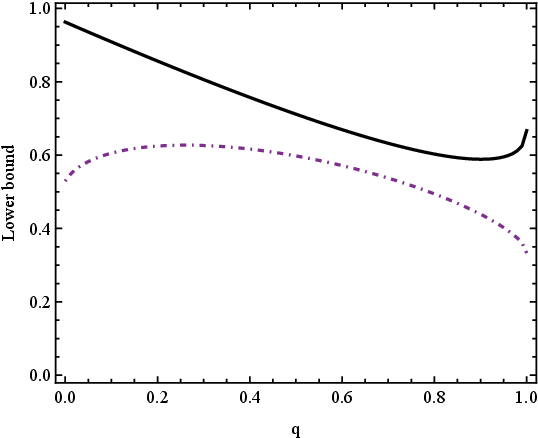}
\end{minipage}}
\caption{{ The black (solid) curve represents the value of the product $|V_{\rho}|(\Phi_{AD})|V_{\rho}|(\Phi_{BF})$ in (a) and (b), or the sum
$|V_{\rho}|(\Phi_{AD})+|V_{\rho}|(\Phi_{BF})$ in (c) and (d). The red (dotted) and the purple (dot-dashed) curves represent the lower bounds in Theorem 1 and 2, respectively. We compare the lower bounds of Theorem 1 with the value of the product $|V_{\rho}|(\Phi_{AD})|V_{\rho}|(\Phi_{BF})$ in (a) for $q=0.2$ and in (b) for $\theta=\pi/3$. The comparison between the lower bound of Theorem 2 and the value of the sum $|V_{\rho}|(\Phi_{AD})+|V_{\rho}|(\Phi_{BF})$ are presented in (c) for $q=0.8$ and (d) for $\theta=3\pi/5$.
\label{fig:Fig1}}}
\end{figure}

{\bf Example 1} Consider the mixed state $\rho=\frac{1}{2}(I_2+\vec{r}\cdot \vec{\sigma})$ with $\vec{r}=(\frac{\sqrt{3}}{3}\cos\theta, \frac{\sqrt{3}}{3}\sin\theta,0)$ and $\sigma_x$, $\sigma_y$, $\sigma_z$ the standard Pauli matrices. We consider the following two quantum channels: the amplitude damping channel $\Phi_{AD}$,
\begin{align*}
\Phi_{AD}(\rho)=\sum_{i=1}^2A_i\rho A_i^\dag, \quad
A_1=|0\rangle\langle0|+\sqrt{1-q}|1\rangle\langle1|, \quad A_2=\sqrt{q}|0\rangle\langle1|;
\end{align*}
and the bit flip channel $\Phi_{BF}$,
\begin{align*}
\Phi_{BF}(\rho)=\sum_{i=1}^2B_i\rho B_i^\dag,\quad  B_1=\sqrt{q}(|0\rangle\langle0|+|1\rangle\langle1|), \quad B_2=\sqrt{1-q}(|0\rangle\langle1|+|1\rangle\langle0|),
\end{align*}
where $0\leq q\leq1$. Fig.~\ref{fig:Fig1} illustrates the lower bounds of the above Theorems, where Fig 1(a) $q=0.2$ and Fig 1(b) $\theta=\pi/3$ show the values of $|V_{\rho}|(\Phi_{AD})|V_{\rho}|(\Phi_{BF})$ and the lower bounds of Theorem 1, Fig 1 (c) $q=0.8$ and Fig 1 (d) $\theta=3\pi/5$ show the lower bound of Theorem 2 and the value of $|V_{\rho}|(\Phi_{AD})+|V_{\rho}|(\Phi_{BF})$. Obviously, the lower bound of Theorem 1 and Theorem 2 are very close to the value of the $|V_{\rho}|(\Phi_{AD})|V_{\rho}|(\Phi_{BF})$ and $|V_{\rho}|(\Phi_{AD})+|V_{\rho}|(\Phi_{BF})$, respectively.

\noindent {\bf 3. Uncertainty relations for arbitrary $N$ quantum channels}

We now provide the summation form of the uncertainty relations based on the $\rho$-absolute variance for arbitrary $N$ quantum channels.

{\bf Theorem 3} Let $\Phi_{1},\cdots,\Phi_N$ be $N$ quantum
channels with Kraus representations
$\Phi_t(\rho)=\sum_{i=1}^{n}K_{i}^{t}\rho (K_{i}^{t})^\dag, ~t=1,2,\cdots,N$ ($N>2$). We have
\begin{align}\label{eq12}
\sum_{t=1}^{N}|V_{\rho}|(\Phi_t)\geq \mathop{\mathrm{max}}\{\overline{LB}1,\overline{LB}2,\overline{LB}3\},
\end{align}
where
\begin{align}\label{eq13}
\overline{LB}1
&=\mathop{\mathrm{max}}\limits_{\pi_t,\pi_s\in S_n}\frac{1}{M N+(N-2)L}\left\{\frac{2L}{N(N-1)}\left[\sum_{i=1}^{n}\left(\sum_{1\leq t<s\leq N}\sqrt{|V_{\rho}|(K_{\pi_{t}(i)}^{t}
+K_{\pi_{s}(i)}^{s})}\right)^{2}\right]\right.
\nonumber\\
&\left.+M\sum_{i=1}^{n}\sum_{1\leq t<s\leq N}|V_{\rho}|(K_{\pi_{t}(i)}^{t}-K_{\pi_{s}(i)}^{s}) +
(M-L)\sum_{i=1}^{n}|V_{\rho}|
\left(\sum_{t=1}^{N}K_{\pi_{t}(i)}^{t}\right)\right\},
\end{align}
\begin{align}\label{eq14}
\overline{LB}2
&=\mathop{\mathrm{max}}\limits_{\pi_t,\pi_s\in S_n}\frac{1}{M N+(N-2)L}\left\{\frac{2M}{N(N-1)}\left[\sum_{i=1}^{n}\left(\sum_{1\leq t<s\leq N}\sqrt{|V_{\rho}|(K_{\pi_{t}(i)}^{t}
-K_{\pi_{s}(i)}^{s})}\right)^{2}\right]\right.
\nonumber\\
&\left.+L\sum_{i=1}^{n}\sum_{1\leq t<s\leq N}|V_{\rho}|(K_{\pi_{t}(i)}^{t}+K_{\pi_{s}(i)}^{s}) +(M-L)\sum_{i=1}^{n}|V_{\rho}|
\left(\sum_{t=1}^{N}K_{\pi_{t}(i)}^{t}\right)\right\},
\end{align}
\begin{align}\label{eq15}
\overline{LB}3
&=\mathop{\mathrm{max}}\limits_{\pi_t,\pi_s\in S_n}\frac{1}{M N+(N-2)L}\left\{\frac{M-L}{(N-1)^2}\left[\sum_{i=1}^{n}\left(\sum_{1\leq t<s\leq N}\sqrt{|V_{\rho}|(K_{\pi_{t}(i)}^{t}
+K_{\pi_{s}(i)}^{s})}\right)^{2}\right]\right.
\nonumber\\
&\left.+L\sum_{i=1}^{n}\sum_{1\leq t<s\leq N}|V_{\rho}|(K_{\pi_{t}(i)}^{t}+K_{\pi_{s}(i)}^{s}) +
M\sum_{i=1}^{n}\sum_{1\leq t<s\leq N}|V_{\rho}|(K_{\pi_{t}(i)}^{t}-K_{\pi_{s}(i)}^{s}) \right\},
\end{align}
with $S_n$ the $n$-element permutation group and $\pi_{t},\pi_{s}\in S_n$ arbitrary $n$-element permutations.

{\it Proof} We use the following norm inequalities presented in Appendix B of \cite{HLTG},
\begin{align}\label{eq16}
\sum_{t=1}^{N} \| u_t\|^2
&\geq\frac{1}{MN+(N-2)L}\left\{\frac{2L}{N(N-1)}\left(\sum_{1\leq t<s\leq N}\| u_t+u_s\|\right)^2+\right.
\nonumber\\
&\left.M\sum_{1\leq t<s\leq N}\| u_t-u_s\|^2+
(M-L)\left\|\sum_{t=1}^{N}u_t\right\|^2\right\}
\end{align}
for $M\geq L>0$,
\begin{align}\label{eq17}
\sum_{t=1}^{N} \|u_t\|^2
&\geq\frac{1}{MN+(N-2)L}\left\{\frac{2M}{N(N-1)}\left(\sum_{1\leq t<s\leq N}\| u_t-u_s\|\right)^2+\right.
\nonumber\\
&\left.L\sum_{1\leq t<s\leq N}\| u_t+u_s\|^2+
(M-L)\left\|\sum_{t=1}^{N}u_t\right\|^2\right\}
\end{align}
for $L\geq M>0$, and
\begin{align}\label{eq18}
\sum_{t=1}^{N} \| u_t\|^2
&\geq\frac{1}{MN+(N-2)L}\left\{\frac{M-L}{(N-1)^2}\left(\sum_{1\leq t<s\leq N}\| u_t+u_s\|\right)^2+\right.
\nonumber\\
&\left.L\sum_{1\leq t<s\leq N}\| u_t+u_s\|^2+M\sum_{1\leq t<s\leq N}\| u_t-u_s\|^2
\right\}
\end{align}
for $L>M>0$. By substituting $u_t$ and $u_s$ with $\widetilde{K}^t_{{\pi_t}(i)}$ and $\widetilde{K}^s_{{\pi_s}(i)}$, respectively,
we obtain the Theorem $3$. $\Box$


The uncertainty quantification of quantum channel $\Phi$ based on the $\rho$-absolute variance can also be expressed in the following form,
\begin{align}\label{eq19}
|V_{\rho}|(\Phi)=\mathrm{Tr}(\alpha^\dag \alpha)=\|\alpha\|^2,
\end{align}
where $\alpha^\dag=(\widetilde{K}^\dag_1,\widetilde{K}^\dag_2,\cdots,\widetilde{K}^\dag_n)$. By using the inequalities $(\ref{eq16})$-$(\ref{eq18})$ and equation $(\ref{eq19})$, we have the following uncertainty relation.

{\bf Theorem 4} Let $\Phi_{1},\cdots,\Phi_N$ be $N$ quantum
channels with Kraus representations
$\Phi_t(\rho)=\sum_{i=1}^{n}K_{i}^{t}\rho (K_{i}^{t})^\dag, ~t=1,2,\cdots,N$ ($N>2$). We have
\begin{align}\label{eq20}
\sum_{t=1}^{N}|V_{\rho}|(\Phi_t)\geq \mathop{\mathrm{max}}\{\widetilde{LB}1,\widetilde{LB}2,\widetilde{LB}3\},
\end{align}
where
\begin{align}\label{eq21}
\widetilde{LB}1
&=\mathop{\mathrm{max}}\limits_{\pi_t,\pi_s\in S_n}\frac{1}{MN+(N-2)L}\left\{\frac{2L}{N(N-1)}\left[\sum_{1\leq t<s\leq N}\sqrt{\sum_{i=1}^{n}|V_{\rho}|(K_{\pi_{t}(i)}^{t}
+K_{\pi_{s}(i)}^{s})}\right]^{2}\right.
\nonumber\\
&\left.+M\sum_{1\leq t<s\leq N}\sum_{i=1}^{n}|V_{\rho}|(K_{\pi_{t}(i)}^{t}-K_{\pi_{s}(i)}^{s}) +
(M-L)\sum_{i=1}^{n}|V_{\rho}|
\left(\sum_{t=1}^{N}K_{\pi_{t}(i)}^{t}\right)\right\},
\end{align}
\begin{align}\label{eq22}
\widetilde{LB}2
&=\mathop{\mathrm{max}}\limits_{\pi_t,\pi_s\in S_n}\frac{1}{MN+(N-2)L}\left\{\frac{2M}{N(N-1)}\left[\sum_{1\leq t<s\leq N}\sqrt{\sum_{i=1}^{n}|V_{\rho}|(K_{\pi_{t}(i)}^{t}
-K_{\pi_{s}(i)}^{s})}\right]^{2}\right.
\nonumber\\
&\left.+L\sum_{1\leq t<s\leq N}\sum_{i=1}^{n}|V_{\rho}|(K_{\pi_{t}(i)}^{t}+K_{\pi_{s}(i)}^{s}) +
(M-L)\sum_{i=1}^{n}|V_{\rho}|
\left(\sum_{t=1}^{N}K_{\pi_{t}(i)}^{t}\right)\right\},
\end{align}
\begin{align}\label{eq23}
\widetilde{LB}3
&=\mathop{\mathrm{max}}\limits_{\pi_t,\pi_s\in S_n}\frac{1}{MN+(N-2)L}\left\{\frac{M-L}{(N-1)^2}\left[\sum_{1\leq t<s\leq N}\sqrt{\sum_{i=1}^{n}|V_{\rho}|(K_{\pi_{t}(i)}^{t}
+K_{\pi_{s}(i)}^{s})}\right]^{2}\right.
\nonumber\\
&\left.+L\sum_{1\leq t<s\leq N}\sum_{i=1}^{n}|V_{\rho}|(K_{\pi_{t}(i)}^{t}+K_{\pi_{s}(i)}^{s}) +
M\sum_{1\leq t<s\leq N}\sum_{i=1}^{n}|V_{\rho}|(K_{\pi_{t}(i)}^{t}-K_{\pi_{s}(i)}^{s}) \right\},
\end{align} where $\pi_{t},\pi_{s}\in S_n$ are arbitrary $n$-element permutations. The parameters $L$, $M$ in $\widetilde{LB}1$, $\widetilde{LB}2$ and $\widetilde{LB}3$ satisfy $M\geq L>0$, $L\geq M>0$ and $L>M>0$, respectively.

Theorem 3 and Theorem 4 characterize the uncertainty relations in different ways. They are not equal in general. In the Appendix D of Ref. \cite{HLTG}, the authors gave an optimal lower bound by comparing two kinds of lower bounds in terms of the metric-adjusted skew information. Here, by replacing the metric-adjusted skew information with the $\rho$-absolute variance, we have the following similar conclusion,
\begin{align}\label{eq24}
\sum_{t=1}^{N}|V_{\rho}|(\Phi_t)\geq \mathop{\mathrm{max}}\{\widetilde{LB}1,\widetilde{LB}2,\overline{LB}3\},
\end{align}
where $\overline{LB}3$ is given in Theorem 3. For convenience we denote by $\widetilde{LB}$ the right hand side of inequality $(\ref{eq24})$. To illustrate our results, we take $M=2,L=1$ for $\widetilde{LB}1$, and $M=1,L=2$ for $\widetilde{LB}2$ and $\overline{LB}3$ in inequality $(\ref{eq24})$.

\begin{figure}[ht]\centering
\subfigure[]
{\begin{minipage}[Xu-Cong-uncertainty-6-2a]{0.49\linewidth}
\includegraphics[width=0.98\textwidth]{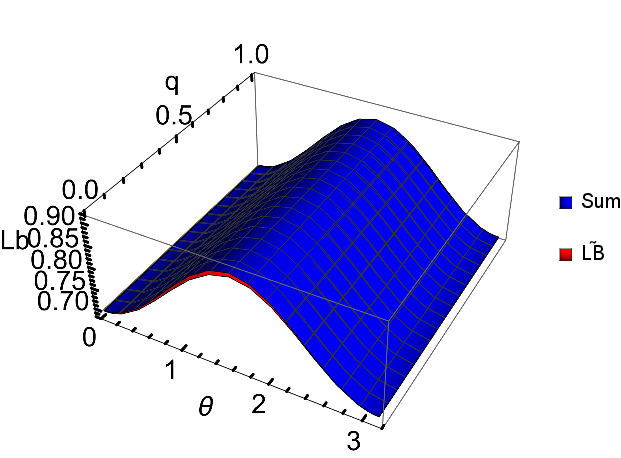}
\end{minipage}}
\subfigure[]
{\begin{minipage}[Xu-Cong-uncertainty-6-2b]{0.49\linewidth}
\includegraphics[width=0.98\textwidth]{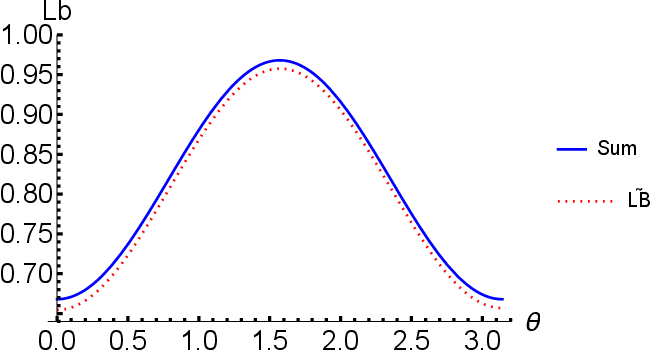}
\end{minipage}}
\caption{{The blue and red surfaces in (a) for $0\leq q\leq1$, and the blue and dotted red curves in (b) for $q=0.1$ represent the sum $V_{\rho}(\Phi_{PD})+V_{\rho}(\Phi_{BF})+V_{\rho}(\Phi_{AD})$ and the lower bounds of inequality $(\ref{eq24})$, respectively.
\label{fig:Fig2}}}
\end{figure}

{\bf Example 2} We consider the quantum state in Example 1 and the following three quantum channels:
(i) the amplitude damping channel $\Phi_{AD}$, (ii) the bit flip channel $\Phi_{BF}$ and (iii) the phase damping channel $\Phi_{PD}$,
\begin{align*}
\Phi_{PD}(\rho)=\sum_{i=1}^2C_i\rho C_i^\dag,\quad  C_1=|0\rangle\langle0|+\sqrt{1-q}|1\rangle\langle1|, \quad C_2=\sqrt{q}|1\rangle\langle1|
\end{align*}
with $0\leq q\leq1$. Fig.~\ref{fig:Fig2} shows the value of $V_{\rho}(\Phi_{PD})+V_{\rho}(\Phi_{BF})+V_{\rho}(\Phi_{AD})$ and the lower bound of $\widetilde{LB}$. Obviously, the blue surface covers fully the red one for arbitrary $0\leq q\leq1$. In particular, we take $q=0.1$ in Fig. 2(b).

\vskip0.1in
\noindent {\bf 4. Conclusions}\\\hspace*{\fill}\\
By recalling the concept of the $\rho$-absolute variance \cite{Gudder}, we have introduced the uncertainty of a quantum channel $\Phi$ in terms of the $\rho$-absolute variance. The elegant properties of $V_{\rho}(\Phi)$ have been explored. In addition, by using the Cauchy-Schwarz inequality and the parallelogram law, we have established the summation and product forms of the uncertainty relations for two quantum channels. By using some fundamental norm inequalities, we have also investigated the summation form of the uncertainty relations for arbitrary $N$ quantum channels. An optimal lower bound have been obtained. Moreover, we have presented several specific examples to illustrate the lower bounds of our results. Our results may shed some new light on the study of uncertainty relations for quantum channels.
\vskip0.1in

\noindent

\subsubsection*{Acknowledgements}
This work was supported by National Natural Science Foundation of
China (Grant Nos. 12161056, 12075159, 12171044); the
specific research fund of the Innovation Platform for Academicians of Hainan Province under Grant No. YSPTZX202215; and Changsha University of Science
and Technology (Grant No. 097000303923).

\subsubsection*{Conflict of interest}
\small {The authors declare that they have no conflict of interest.}




\begin{thebibliography}{S2}

\bibitem{FCP} Fuchs, C. A., Peres, A.: Quantum-state disturbance versus information gain: uncertainty relations for
quantum information.  Phys. Rev. A \textbf{53}, 2038 (1996)

\bibitem{RJB} Renes, J. M., Boileau, J.C.: Conjectured strong complementary information tradeoff.  Phys. Rev. Lett.
\textbf{103}, 020402 (2009)

\bibitem{BSL} Bowen, W. P., Schnabel, R., Lam, P.K., Ralph, T.C.:  Experimental investigation of criteria for continuous
variable entanglement.  Phys. Rev. Lett. \textbf{90}, 043601 (2003)

\bibitem{GO} G\"{u}hne, O,.:  Characterizing entanglement via uncertainty relations.  Phys. Rev. Lett. \textbf{92}, 117903 (2004)

\bibitem{HJBR} Howell, J. C., Bennink, R. S., Bentley, S. J., Boyd, R. W.:  Realization of the einstein-podolsky-rosen paradox using momentum-and position-entangled photons from spontaneous parametric down conversion.
 Phys. Rev. Lett. \textbf{92}, 210403 (2004)


\bibitem{OW} Oppenheim, J., Wehner, S.: The uncertainty principle determines the nonlocality of quantum mechanics.  Science
\textbf{330}, 1072 (2010)





\bibitem{HW} Heisenberg, W.:  \"Uber den anschaulichen Inhalt der
quantentheoretischen Kinematik und Mechanik.  Z. Phys.
\textbf{43}, 172 (1927)


\bibitem{KEH} Kennard, E. H.: Zur quantenmechanik einfacher bewegungstypen.  Z. Phys. \textbf{44}, 4 (1927)


\bibitem{WH} Weyl, H.:  Gruppentheorie und Quantenmechanik Hirzel.  Leipzig (1928)


\bibitem{RH} Robertson, H. P.: The uncertainty principle.  Phys. Rev. \textbf{34}, 163 (1929)

\bibitem{SE} Schr\"{o}dinger, E.:  Zum Heisenbergschen Unsch\"{a}rfeprinzip. Sitzungsber.  Preuss. Akad. Wiss. Phys. Math. KI. \textbf{14}, 296–303 (1930)

\bibitem{MP} Maccone, L., Pati, A. K.:   Stronger uncertainty relations for all incompatiable observables.  Phys. Rev. Lett. \textbf{113}, 260401 (2014)

\bibitem{WZB} Wang, K., Zhan, X., Bian, Z., Li, J., Zhang, Y., Xue, P.:    Experimental investigation of the stronger uncertainty relations for all incompatible observables. Phys. Rev. A \textbf{93}, 052108 (2016)


%
%
%
%


\bibitem{LS1} Luo, S.: Heisenberg uncertainty relation for mixed states. Phys. Rev. A \textbf{72}, 042110 (2005)
\bibitem{LS2} Luo, S.: Quantum versus classical uncertainty.  Theor. Math. Phys. \textbf{143}, 681 (2005)
\bibitem{LS3} Luo, S.: Quantum uncertainty of mixed states based on skew information.  Phys. Rev. A \textbf{73}, 022324 (2006)






\bibitem{Gudder} Gudder, S.: Operator probability theory.  Int. J. Pure Appl. Math. \textbf{39}, 511 (2007)



\bibitem{DD1}Dou, Y., Du, H.:  Generalizations of the Heisenberg and Schr\"{o}dinger uncertainty relations.  J. Math. Phys. \textbf{54}, 103508 (2013)

 \bibitem{SL} Sun, Y., Li, N.: The uncertainty of quantum channels in terms of variance.  Quantum Inf. Process. \textbf{20}, 25 (2021)





\bibitem{Wu1} Wu, Z., Zhang, L., Wang, J., Li-Jost, X., Fei, S.-M.:  Uncertainty relations based on modified Wigner-Yanase-Dyson skew information.  Int. J. Theor. Phys. \textbf{59}, 704 (2020)



\bibitem{LS}Luo, S., Sun, Y.: Coherence and complementary in state-channel interaction. Phys. Rev. A \textbf{98}, 012113 (2018)


\bibitem{XWF3} Xu, C., Wu, Z., Fei, S.-M.:  Uncertainty of quantum channels via modified generalized variance and modified generalized Wigner-Yanase-Dyson
    skew information.  Quantum Inf. Process. \textbf{21}, 292 (2022)



\bibitem{ZN} Zhou, N., Zhao, M., Wan, Z. G. Li, T.: The uncertainty relation for quantum channels based on skew information.  Quantum Inform. Process. \textbf{22}, 6 (2023)

\bibitem{ZL} Zhang, L., Gao, T., Yan, F.:  Tighter uncertainty relations based on Wigner-Yanase skew information for observables and channels.  Phys. Lett. A \textbf{387}, 127029 (2021)




%








\bibitem{HJ1}H, X., J, N.: Enhanced quantum channel uncertainty relations by skew information.  Quantum Inf. Process. \textbf{22}, 365 (2023)

\bibitem{FSS} Fu, S., Sun, Y., Luo, S.:  Skew information-based uncertainty relations for quantum channels.  Quantum Inf. Process. \textbf{18}, 258 (2019)



\bibitem{ZWF1} Zhang, Q. H., Wu, J. F., Fei, S.-M.: A note on uncertainty relations of arbitrary $N$ quantum channels.  Laser Phys. Lett. \textbf{18}, 095204 (2021)

\bibitem{ZWF2} Zhang, Q. H., Wu, J. F., Ma, X., Fei, S.-M.: A note on uncertainty relations of metric-adjusted skew information. Quantum Inf. Process. \textbf{22}, 115 (2023)

\bibitem{ZWF5} Zhang, Q. H., Wu, J., Fei, S.-M.:  A note on uncertainty relations of arbitrary $N$ quantum channels.  Laser Phys. Lett.
\textbf{18}, 095204 (2021)

\bibitem{ZWF7}Zhang, Q. H., Fei, S.-M.:  Wigner-Yanase skew information-based uncertainty relations for
quantum channels. Eur. Phys. J. Plus \textbf{139}, 137 (2024)

\bibitem{HLTG} Li, H., Gao, T., Yan, F.:  Tighter sum uncertainty relations via metric-adjusted skew information.  Phys. Scr. \textbf{98}, 015024 (2023)

\bibitem{CAL} Cai, L.: Sum uncertainty relations based on metric-adjusted skew information. Quantum Inf. Process. \textbf{20}, 72 (2021)



\bibitem{RRNL} Ren, R., Li, P., Ye, M., Li, Y.: Tighter sum uncertainty relations based on metric-adjusted skew information.  Phys. Rev. A
\textbf{104}, 052414 (2021)




\bibitem{XWF1} Xu, C., Wu, Z., Fei, S.-M.: Sum uncertainty relations based on $(\alpha,\beta,\gamma)$ weighted Wigner-Yanase-Dyson skew information. Int. J. Theor. Phys. \textbf{61}, 185 (2022)

\bibitem{XWF2} Xu, C., Wu, Z., Fei, S.-M.: Tighter uncertainty relations based on $(\alpha,\beta,\gamma)$ modified weighted Wigner-Yanase-Dyson skew information of quantum channels.  Laser Phys. Lett. \textbf{19}, 105206 (2022)

\bibitem{XWF4} Xu, C., Wu, Z., Fei, S.-M.: Tighter sum uncertainty relations via $(\alpha,\beta,\gamma)$ weighted Wigner-Yanase-Dyson skew information.  Commun. Theor. Phys. \textbf{76}, 035102 (2024)

\bibitem{NC} Nielson, M.A., Chuang, I.L.:  Quanutm Computation and Quantum Information  Cambridge University Press, Cambridge (2011)


\bibitem{BHATIA} Bhatia, R.: Partial traces and entropy inequalities. Linear Algebra Appl. \textbf{370}, 125 (2003)








\bibitem{YNB} Xiao, Y., Jing, N., Yu, B., Fei, S.-M., Li-Jost, X.:  Near-optimal variance-based
uncertainty relations.  Front. Phys. Lett. \textbf{10}, 846330 (2022)




%
%








%
%
%
%
%
%
%
%
%
%
%
%
%
%
%
%
%
%




\end{thebibliography}
\end{document}